# REJUVENATION OF THE CONTROLS OF THE CERN ISOLDE FACILITY USING INDUSTRIAL COMPONENTS

J. Allard, O. Jonsson, F. Locci, G. Mornacchi, CERN, Geneva, Switzerland

Abstract

In the context of the general consolidation of the CERN ISOLDE facility, a project has been started to upgrade the ISOLDE control system. We describe the new ISOLDE control system, emphasizing the systematic use of industrial components such as PLCs and field buses, their integration with the existing, VME based, CERN PS control system and their potential applicability to both existing and new controls problems at the CERN PS complex. We also discuss how to extend a PLC-based solution to the case where real-time response is an issue.

## 1 INTRODUCTION

The On-Line mass-separator (ISOLDE) is a facility for the production of pure low energy radioactive ion beams at CERN [1]. A project for its consolidation has been started and the upgrade of the control system has been identified as one of the urgent jobs. To this end a sub-project has been defined with the mandate to consolidate the equipment layer of the ISOLDE control system and to integrate this latter with the PS controls infrastructure on the time scale of the ISOLDE 2002 start-up. The project builds on a previous study for the rejuvenation of the ISOLDE control system [2].

In this paper we describe the solution devised for the ISOLDE control system and report on the status of its implementation. Emphasis is put on the application of industrial components.

## 2 THE ISOLDE CONTROL SYSTEM

### 2.1 The existing ISOLDE control system

The existing ISOLDE control system [3] is based on commodity hardware (PCs), using Microsoft operating systems (DOS and WindowsNT) and tools (e.g. Access, Excel). The system is layered according to the accelerator controls "standard model": a supervisory layer, consisting of PCs running console application software, and a front-end (or equipment control) layer, also based on PCs, which interfaces to the equipment hardware and runs the equipment specific control software. The integration of the different (about 30) PCs is realised with a in-house, high level communication protocol over TCP/IP and Ethernet. The interface to the equipment (e.g. power supplies) is not homogeneous but is a combination of old standards, such as CAMAC, special PC I/O cards and other ad hoc solutions. A detailed description of the ISOLDE control system is available in [4]. Without discussing its technical merits, the old ISOLDE control system is a unique implementation (no element in common with the PS control system). It is also technologically obsolete and its exploitation and maintainability are a very serious concern. Hence two key features of the design of the new ISOLDE controls: compatibility (with the PS control system) and use of industrial components (reliable, off-the-shelf technology).

### 2.2 The new front-end layer

A complete description of the new ISOLDE front-end controls is available in [5]. For some ISOLDE equipment, in particular the beam instrumentation, the equipment control layer will be implemented using the standard PS controls solution: the Device Stub Controller (DSC) [6]. This is a VME based system including a processor, running the LynxOS operating system, and various VME interface modules for the I/O towards the equipment. It is well adapted to buses such as GPIB or the MIL1553 standard.

For the rest of the equipment we have devised a solution based on the conclusions of a previous study [2], which systematically applies industrial components. The main ingredients are:

- The Device Stub Controller (DSC): this acts as a communication concentrator for the PLCs for equipment whose control is implemented with industrial components. The DSC connects a PLC based sub-system to the PS control system by translating the equipment control requests (from the console layer) into a simple control protocol (towards the PLC layer). The DSC glues together different environments: the supervisory layer, the service infrastructure (e.g. the databases) and the control process.
- The Programmable Logic Controller (PLC): the task of the control process is delegated to a PLC. One PLC is dedicated to a number of instances of equipment of the same family (such as a particular type of a power supply), and implements the local control process as well as maintenance functions. A dedicated PC or a simple display module may be installed to operate the PLC locally, when offline from the control system supervisory layer.
- PROFIBUS [7]: this is the field bus for the standard industrial I/O modules (e.g. ADC, DAC, etc.) which connects to the equipment on either copper wires or an optical fibre (for example in the case of equipment running in a high voltage environment). While this is not the case at ISOLDE, a PLC controlling some complex equipment could delegate part of the control to a hierarchy of PLCs on PROFIBUS.

- Ethernet: it is used as the fieldnetwork to connect the DSC to the PLCs. This choice greatly simplifies the connection of our VME processor to the PLC. It allows us to benefit from the general CERN network infrastructure and support, and it opens the door to the use of e.g. Web-like applications. It also lifts the constraints on the frame size (i.e. the PROFIBUS limitation to 256 bytes/frame) and the communication paradigm (i.e. the PROFIBUS mailbox-like protocol).
- Protocols: given the selection of Siemens hardware, we run the RFC1006 [8] protocol on top of the Ethernet fieldnetwork. We also run the Siemens Fetch/Write protocol [9] to remotely execute load/store operations in the PLC memory.

In summary, the new control system of the ISOLDE facility includes two configurations for the equipment control layer: one is the standard, VME based DSC, while in the second, PLCs with industrial I/O modules on PROFIBUS are systematically used. In this latter the control process has been migrated from the DSC to the PLC; the former plays the role of link to the rest of the control system and its facilities. The main ingredients defined above are put together in figure 1, which depicts the layout for ISOLDE equipment controlled via industrial components.

We have designed and developed a simple client-server protocol, on top of RFC1006, to implement the generic communication interface between the DSC and the PLC.

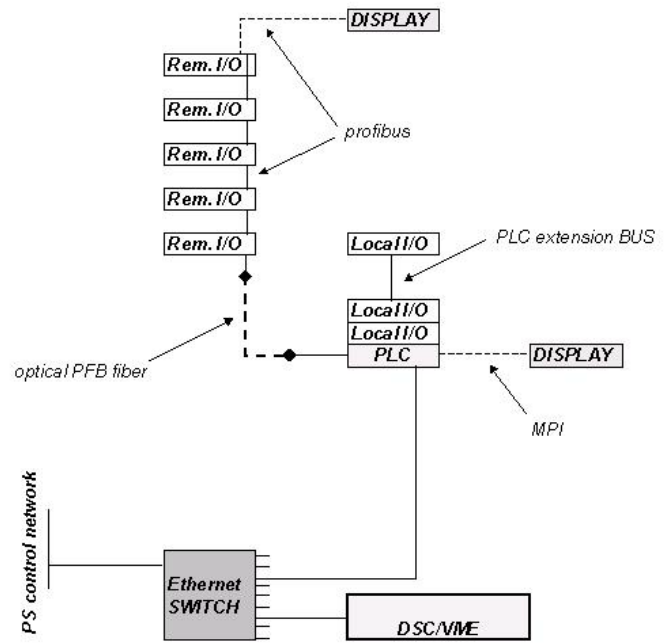

Figure 1: Generic PLC based control layout.

The PLC software includes a small generic component that serves requests from the DSC and activates the relevant equipment specific threads of code. In the spirit of keeping the PLC programs as simple as possible, the state of the equipment is maintained by the DSC, so as to exploit all the facilities (such as dynamic data tables) which are available in the PS control system. A pictorial view of the organisation is shown in figure 2.

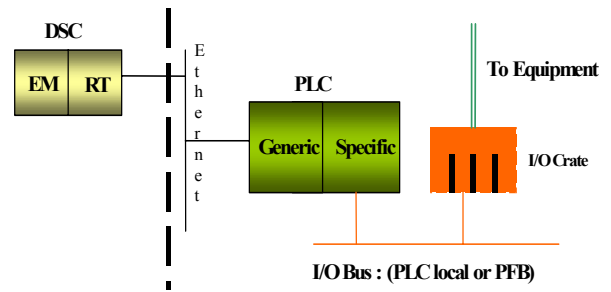

Figure 2: Organization of the equipment control layer.

The sharp boundary between the generic (communication) part and the equipment specific control process greatly simplifies both the task of the software developer and the communication protocol. This latter is reduced to the exchange of a data structure.

CERN recommends a number of industrial components standards, products and manufacturers. We have chosen PROFIBUS as the I/O field bus and Siemens products for both PLCs and I/O modules. Depending on the required performance and I/O channel density, we have used either the SIMATIC S7-300 or the SIMATIC S7-400 series PLCs and I/O modules. The PLC software is developed using the Pascal-like SCL language. The choice of one brand does not prevent the use of other industrial components, however. Field buses and field networks allow different components to inter-operate, indeed we foresee to complement our architecture with other CERN recommended products.

As an example, take the system for the control of ~300 power supplies for the ISOLDE electrostatic lenses. They will be controlled by a SIMATIC S7-400 PLC based system. We currently estimate that a single S7 414-2 PLC will be enough to control the ~300 power supplies. Industrial I/O modules (ADC, DAC etc.) are accommodated in 3 segments of 4 crates each using the PLC extension bus. Equipment specialists will use a local display module for local configuration and diagnostics as well as to perform local control.

The status of the developments is well advanced: a few sub-systems will be installed at ISOLDE during fall, and if successful left in operation. The bulk of the new front-end layer will be installed during the 2001-2002 winter shutdown.

## 3 CONCLUSIONS

We have systematically used, where applicable, industrial components in the design of the new ISOLDE control system. The benefits of such an approach are extensively discussed in [2], and include flexibility in the design and development, easy assembly of off-the-shelf components, and tracking the evolution of industry. The approach is well adapted to a wide range of equipment: the power supplies, the stepping motors, the gas system and the vacuum system. For some equipment, in particular where the acquisition of a sizeable amount of data is concerned or the equipment is interfaced via the GPIB bus, we have retained the standard PS controls solution.

We have merged a generic solution based on industrial components with the existing PS control system: on the one hand we made use of the powerful tools already available, while on the other hand we have provided an additional building block to the PS control system. Indeed the solution we have devised is not limited to ISOLDE. The door is open for its use elsewhere at the PS complex, in particular for applications where the PS pulse-to-pulse modulation [6] is not an issue and to the controls of future CERN projects such as the LEIR and CTF3 machines.

The installation of Ethernet as a fieldnetwork was easy and it is very well adapted to a horizontal (device oriented) control system (as opposed to a vertical, channel oriented, system where the fieldbus is better suited). It is however more expensive, the cost of the Ethernet being of the same order as the cost of the PLC.

Finally there are two open questions, both related to the use of Ethernet as a fieldnetwork. The first concerns security: how to regulate the external access to the PLC sub-system, in particular if user interfaces based on an embedded Web server are used. The second is related to the use of a PLC sub-system in the PS environment; here the issue is deterministic (real-time) behavior, the order of magnitude being a few tens of milliseconds. While the PLC per se is deterministic[1], the same is not true for the delivery of e.g. timing signals over Ethernet. We propose to overcome the problem by using a structured deterministic (i.e. collision-less) Ethernet, an issue where industry is currently very active.

## 4 ACKNOWLEDGEMENTS

It is a pleasure to acknowledge the support of the controls group of the CERN PS division, in particular its leader B. Frammery, as well as colleagues from the ISOLDE equipment groups: PS/PO, PS/OP, PS/BD and LHC/VAC.

## REFERENCES

[1] E. Kugler. The ISOLDE facility. Hyperfine Interaction 129 (2000) 23-42.
[2] O. Jonsson, F. Locci, G. Mornacchi. "ISOLDE Front-End Study, Final Report". CERN int. PS/CO Note 2001-009 (Tech.).
[3] O.C. Jonsson et al. "The evolution of the ISOLDE control system". NIM B, 126 (1997) 30-34.
[4] F. Locci. Le système de contrôle ISOLDE et REX". CERN int. PS/CO Note 99-28.
[5] J. Allard, O. Jonsson, F. Locci, G. Mornacchi. "Consolidation of the ISOLDE controls". CERN int. PS/CO Note, to be published.
[6] http://wwwpsco.cern.ch/.
[7] http://www.profibus.com.
[8] http://www.faqs.org/rfcs/rfc1006.html.
[9] SIMATIC CP 1430 TCP Coupler. Reference Manual. Siemens AG.

---

[1] Provided the PLC is capable of executing the cycle within the required time frame.